\documentstyle[aps,preprint]{revtex}

\begin{document}
\input epsf

\hfill\vbox{\hbox{\bf CPP-95-17}\hbox{\bf DOE-ER40757-073}
\hbox{Nov 1995}}\\

\begin{center}
{\Large \bf 
Probing the Chromoelectric and Chromomagnetic dipole moments of the Top
quark at Hadronic Colliders}

\vspace{0.2in}

Kingman Cheung\footnote{Internet address: {\tt cheung@utpapa.ph.utexas.edu}}

{\it Center for Particle Physics, University of Texas, Austin, TX 78712}

\end{center}

\begin{abstract}
We study the effects of the anomalous chromoelectric and chromomagnetic
dipole moments of the top quark on the top-pair production, on the top-pair
plus 1jet production, and on their ratios, as well as their dependence on 
the transverse
momenta of the top quark and the jet.  We also construct CP-odd and 
$\hat T$-odd observables to probe the dispersive part of the chromoelectric
form factor of the top quark, and compare their sensitivities in both
top-pair and top-pair plus 1 jet production.
\end{abstract}

\thispagestyle{empty}

\section{Introduction}

The year 1995 with the discovery of the sixth quark -- the top quark 
\cite{cdf,d0} marks the triumph for the standard model (SM).
The future will be an era of looking for physics beyond the SM.   
There have been many theories that are pointing to different directions.  
Only new signs of experimental evidence can point to the right theory. 
After the discovery of the top quark studies of the properties of the top 
quark will be the major goal of the Tevatron in the next 10 years.  
In the next run of the Tevatron both the center of mass energy and the 
luminosity will be upgraded.  The center of mass energy will be at $\sqrt{s}=
2$ TeV and the yearly luminosity could be as much as 1--5 fb$^{-1}$ 
\cite{scott}.
After the Tevatron the Large Hadron Collider (LHC) will further examine the 
properties of the top quark with much higher statistics,  and will begin to 
investigate the ultimate mechanism for the electroweak symmetry breaking.  

The top is so heavy that it provides interesting avenues to new 
physics, via the production \cite{aas,kane,ma,rizzo,hab}, decays 
\cite{gunion}, and their scattering \cite{me}.  One interesting property 
of the top quark is its anomalous chromomagnetic (CMDM) and 
chromoelectric (CEDM) dipole moments, which will affect the production, 
decays, and the interactions of the top quark  with other particles.  
In this work we concentrate on the production channel in a $p\bar p$ collider 
to probe the anomalous dipole moments of the top quark in a model-independent 
way, using an effective Lagrangain approach. 
The effects of the anomalous dipole moments on top quark production at hadronic
colliders have been studied in some details \cite{aas,ma,rizzo,hab}. 
The study of CP-violating effects in top-pair production due to the CEDM 
of the top quark were carried out in Refs.~\cite{aas,ma}.  
However, it was first suggested by Atwood, Kagan, and Rizzo \cite{rizzo}
to use the total cross sections and transverse momentum distributions
of top-pair production to constrain the CMDM of the top quark, and 
later extended in Ref.~\cite{hab} to include the CEDM also.
Here we extend the study to the production of $t\bar t$ pair plus 1 jet 
(denoted by $t\bar tj$), and the ratio of 
$t\bar t j$  to $t\bar t$ production, as well as their 
dependence on the transverse momenta of the top quark and the extra jet.
This is the first purpose of the paper. 
  The advantages of studying the 
ratio include (i) the elimination of systematic uncertainties arising from
experimental detections, higher order corrections, and the factorization 
scale, and (ii) the possibility of studying the dependence on the 
transverse momentum of the extra jet.  In this work we use the 
parameters of the next run at the Tevatron, i.e., $\sqrt{s}=2$ TeV and
yearly luminosity of order of a few fb$^{-1}$ \cite{scott}, and throughout
the paper we choose the top mass $m_t=176$ GeV \cite{cdf}.

The measurement of the electric dipole moments of neutrons and electrons 
has been a major effort in pursuing CP-violation other than in the kaon 
system because a nonzero electric dipole moment is a clean 
signal of CP-violation.   
Since the top quark is heavy, it can have potentially large couplings to
the Higgs bosons of the underlying theory (e.g. multi-Higgs doublet model
\cite{wein}) that naturally contains CP-violating phases other than the 
usual CKM phase, and, therefore, the CEDM of the top quark can be 
potentially large.
Thus, probing the CEDM of the top quark is important in searching for 
CP-violation, and it requires
a CP-odd observable.  CP-odd observables can further be classified by their 
$\hat T$ behavior, where $\hat T$ is the naive time reversal that transforms
the kinematic observables according to the time reversal ($t\to -t$) but 
without interchanging the initial and final states.  A CP-odd and $\hat T$-even
observable can probe the imaginary part of the CEDM form factor of the top
quark, while a CP-odd and $\hat T$-odd observable can probe the dispersive
part \cite{darwin}.   
The simplest example of CP-odd and $\hat T$-odd observables will be
a triple product of 3-momenta.
Since at $p\bar p$ machines the initial state is a CP eigenstate, a 
non-zero expectation value for a CP-odd observable is an
indication of CP-violation.
The second purpose of the paper is to examine the expectation values for 
some CP-odd and $\hat T$-odd observables in $t\bar t$ and $t\bar t j$
production, and compare their sensitivities.

The organization of the paper is as follows.  In Sec. II we write down the
effective Lagrangian and detail the calculation method.  In Sec. III we show
the results of the total and differential cross sections for $t\bar t$ 
and $t\bar t j$ production.  In Sec. IV we discuss some CP-odd 
observables that are sensitive to the CEDM of the
top quark.  We shall conclude in Sec. V.  
Detail helicity formulas for the calculations are given in the appendix.
Finally, we also emphasize that a similar
study on the present  data of $b$ quark production should be useful in
constraining the anomalous dipole moments of the $b$ quark.

\section{Effective Lagrangian}

The effective Lagrangian for the interactions between the top quark and gluons
that include the CEDM and CMDM form factors is 
\begin{equation}
\label{eff1}
{\cal L}_{\rm eff} = g_s \bar t T^a \left[ -\gamma^\mu G_\mu^a +
\frac{F_2(q^2)}{4m_t} \sigma^{\mu\nu} G_{\mu\nu}^a - 
\frac{i F_3(q^2)}{4m_t} \sigma^{\mu\nu} \gamma^5 G_{\mu\nu}^a \right ] t \;,
\end{equation}
where $T^a$ is the SU(3) color matrices, $G_{\mu\nu}^a$ is the gluon field
strength, $\sigma^{\mu\nu}=\frac{i}{2}[\gamma^\mu,\gamma^\nu]$, and $F_2(q^2)$
and $F_3(q^2)$ are, respectively, the CMDM and CEDM forms factors of the top
quark.  
Since such CEDM and CMDM interactions are not renormalizable, they must 
originate from some loop exchanges, and they are $q^2$ dependent and can
develop imaginary parts.  But the imaginary parts must vanish at zero
momentum transfer, so the imaginary parts are related to terms of dimension
greater than five in the effective Lagrangian.  Therefore, for our purposes
we only consider these form factors to be real, as displayed
in Eq.~(\ref{eff1}).
Assuming $|q^2| \ll \Lambda$, where $\Lambda$ is the scale of new physics,
the form factors can be approximated by 
\begin{equation}
F_2(q^2) \approx \kappa, \qquad F_3(q^2)\approx \tilde{\kappa} \qquad
{\rm for}\; |q^2| \ll \Lambda \;,
\end{equation}
where $\kappa$ and $\tilde{\kappa}$ are independent of $q^2$.
The CEDM of the top quark is then given by $g_s \tilde{\kappa}/(2m_t)$, 
while CMDM is $g_s \kappa/(2m_t)$.   
For $\tilde \kappa=1$ and $m_t=176$ GeV, the CEDM of the top quark is
about $5.6 \times 10^{-17} g_s \cdot {\rm cm}$.
In terms of $\kappa$ and $\tilde{\kappa}$ 
the effective Lagrangian becomes
\begin{equation}
\label{eff2}
{\cal L}_{\rm eff} = g_s \bar t T^a \left[ -\gamma^\mu G_\mu^a +
\frac{\kappa}{4m_t} \sigma^{\mu\nu} G_{\mu\nu}^a - 
\frac{i \tilde{\kappa}}{4m_t} \sigma^{\mu\nu} \gamma^5 G_{\mu\nu}^a 
\right ] t \;.
\end{equation}
We can then derive the effective interaction of the $ttg$ vertex 
\begin{equation}
\label{ttg}
{\cal L}_{t_i t_j g} = -g_s \bar t_j T^a_{ji} \left[ \gamma^\mu +
\frac{i}{2m_t} \sigma^{\mu\nu} q_\nu (\kappa - i \tilde{\kappa} \gamma^5) 
\right ] t_i\; G_\mu^a \;,
\end{equation}
where $t_i (t_j)$ is the incoming (outgoing) top quark 
and $q_\nu$ is the 4-momentum
of the outgoing gluon.  The Lagrangian in Eq.~(\ref{eff2}) also induces a 
$ttgg$ interaction given by 
\begin{equation}
\label{ttgg}
{\cal L}_{t_i t_jgg} = \frac{ig_s^2}{4m_t} \bar t_j (T^b T^c - T^c T^b)_{ji}
\sigma^{\mu\nu} ( \kappa -i \tilde{\kappa} \gamma^5) t_i G_\mu^b G_\nu^c \;,
\end{equation}
which is absent in the SM.
All the ingredients are now ready for the calculation of the parton level 
cross sections for the production of $t\bar t$ and $t\bar tj$.

\subsection{$t\bar t$ Production}

The parton-level processes, 
\begin{equation}
\begin{array}{rcl}
q \bar q  & \rightarrow &   t \bar t  \\
gg        & \rightarrow &   t \bar t
\end{array}
\end{equation}
including the CEDM and CMDM interactions have been calculated in 
Ref.~\cite{aas,ma,hab}.  
We did an independent calculation and confirmed the analytic results
of Ref.~\cite{hab}. \footnote{In our definition $\tilde\kappa=2\hat {d}'_t$
and $\kappa=-2\hat{\mu}'_t$ for the definition of $\hat{d}'_t$ and
$\hat{\mu}'_t$ in Ref.~\cite{hab}.}
The contributing Feynman diagrams are shown in 
Fig.~\ref{fig1}, where the dressed vertices corresponding to the $ttg$ 
and $ttgg$ interactions of Eq.~(\ref{ttg}) and Eq.~(\ref{ttgg}) are marked 
by a dot.  
For convenience we present the parton-level cross sections here
\begin{eqnarray}
\label{tt1}
\frac{d\hat\sigma_{q\bar q\to t\bar t}}{d\hat t} &=& 
\frac{8\pi\alpha_s^2}{9\hat s^2} \left[
\frac{1}{2} -v +z -\kappa + \frac{1}{4}(\kappa^2 - \tilde{\kappa}^2) 
+ \frac{v}{4z}(\kappa^2 + \tilde{\kappa}^2) \right ] \\
\label{tt2}
\frac{d\hat\sigma_{gg\to t\bar t}}{d\hat t} &=& 
\frac{\pi \alpha_s^2}{12\hat s^2} \biggr [ \left(\frac{4}{v}-9 \right)
\left(\frac{1}{2} -v +2z(1-\frac{z}{v}) - \kappa (1- \frac{\kappa}{2}) \right)
\nonumber \\
&+&  \frac{1}{4} (\kappa^2 + \tilde{\kappa}^2) \left( \frac{7}{z}(1-\kappa)
+\frac{1}{2v}( 1+\frac{5\kappa}{2}) \right ) 
+ \frac{1}{16} (\kappa^2 + \tilde{\kappa}^2)^2 \left( -\frac{1}{z} +
\frac{1}{v} + \frac{4v}{z^2} \right )
\biggr ] \;.
\end{eqnarray}
We shall examine the effects of $\kappa$ and $\tilde \kappa$ on 
$t\bar t$ production in the next section.

\subsection{$t\bar t$ plus 1 jet Production}

The $t\bar tj$ production including the effects of the CEDM and CMDM 
couplings of the top quark is a new calculation.
The reason to extend the $t\bar t$ production to $t\bar tj$ production
is the richer kinematics that can be constructed in the final state.  Also,
the kick-out of a high $p_T$ gluon enables one to probe the $ttg$ vertex in
different phase space region.
The contributing subprocesses are 
\begin{center}
\begin{enumerate}
\item[(i)] $q\bar q \to t\bar t g$
\item[(ii)] $ gg\to t\bar t g$
\item[(iii)] $q(\bar q) g \to t\bar t q(\bar q)$
\end{enumerate}
\end{center}
where $q$ denotes a light parton ($u,d,s,c$, or $b$).  The contributing 
Feynman diagrams for the subprocesses (i) and (ii) are shown in 
Fig.~\ref{fig2} and Fig.~\ref{fig3}, respectively.  The subprocess (iii) 
can be obtained from subprocess (i)
by a crossing of the $\bar q$ in the initial state with the $g$ in the final
state.  Since the number of diagrams is large in this case it is 
convenient to use the helicity-amplitude method to sum and square the 
amplitudes.  We use the helicity-amplitude method of Ref.~\cite{stange}.
The expressions for the helicity amplitudes are given in the appendix.  
We have checked the gauge invariance of the total amplitude and the Lorentz 
invariance of the amplitude squared.
Moreover, by setting $\tilde \kappa=\kappa=0$ in our calculations 
our results agree with the SM results generated by MADGRAPH \cite{mad}, 
and the results of Ref.~\cite{sexton}.  

Before we leave this section, we specify other inputs in our calculations.
We use the parton distribution functions of CTEQ (v.3) \cite{cteq} of which
we chose the leading order fit.  We also used a simple 1-loop formula for
the running coupling constant $\alpha_s$ as follows
\begin{equation}
\alpha_s(\mu) = \frac{\alpha_s(m_Z)}{1+ \frac{33-2n_f}{6\pi} \;
\alpha_s(m_Z) \; \log\left(\frac{\mu}{m_Z} \right ) } \;,
\end{equation}
where $\alpha_s(m_Z)=0.117$ \cite{pdg} and $n_f$ is the number of active
flavors at the scale $\mu$.  The scale used in the parton distribution
functions and the running coupling constant is chosen to be 
$\sqrt{m_t^2 + p_T^2({\rm top})}$.
For $t\bar t$ production we used the expressions in Eqs.~(\ref{tt1}) -- 
(\ref{tt2}) for the subprocess cross sections, while for $t\bar tj$ 
production we used the helicity amplitudes listed in the appendix.  
The decays of the top and anti-top quarks can be included with full spin
correlation using the helicity amplitude method \cite{stange}.  The 
formulas are also given in the appendix.

\section{Results}
\subsection{Total cross sections}

The total cross sections for $t\bar t$ production at the 
$\sqrt{s}=2$ TeV $p\bar p$ collider are shown in Fig.~\ref{fig4}, as a 
contour plot in $\tilde\kappa$ and $\kappa$.   The SM value given by 
$\tilde\kappa=\kappa=0$ is about 5.2 pb.  From Fig.~\ref{fig4} the cross
section is symmetric about $\tilde{\kappa}=0$ because $\tilde{\kappa}$ only 
appears as even powers in the total cross section.  This is easy to understand
because the total cross section is not a CP-violating observable to separate
the CP-violating form factor $\tilde{\kappa}$.  Also, the total cross section
increases when $\tilde{\kappa}$ moves away from zero.  On the other
hand, the cross section is not symmetric about $\kappa=0$, but instead, 
about $\kappa\simeq 0.8$, due to terms linearly 
proportional to $\kappa$ in the total
cross section.  Measurements of the total cross section at the Tevatron 
(Run II) can impose constraints on $(\tilde{\kappa},\kappa)$ plane.  

Figure~\ref{fig5} shows the contours of the cross sections of the 
$t\bar t j$ production with $p_T(j) > 5,10,20$ GeV, respectively, 
in (a), (b), and (c).  
Qualitative behaviors are very similar to that of $t\bar t$ production. 
The SM cross sections with $p_T(j)>5,10,20$ GeV are 4, 2.5, and 1.3 pb,
respectively.  Therefore, measurements of $t\bar tj$  cross sections
can further impose constraints on the $(\tilde{\kappa},{\kappa})$ plane.

The more interesting quantity is the ratio of the two cross sections:
$\sigma(t\bar tj)/\sigma(t\bar t)$ with various $p_T(j)$ cuts.
This ratio is shown in Fig.~\ref{fig6}(a), (b), and (c) for 
$p_T(j) > 5,10,20$ GeV, respectively.   From Fig.~\ref{fig6}(a)--(c) 
we can see that the ratio
is smallest around the region ($\tilde{\kappa}=0, {\kappa}=0$).  
Though the quantitative behaviors of $t\bar tj$ production are very
similar to $t\bar t$ production, there are regions on $(\tilde{\kappa},\kappa)$
plane that the $t\bar tj$ production increases proportionately much
more than $t\bar t$ production, as shown by, e.g.,  the contours with 
ratio greater than 1 in Fig.~\ref{fig6}(a).  This is not unexpected because
of extra dressed $ttg$ and $ttgg$ vertices in $t\bar tj$ production.  
Presumably, if a jet of 5 GeV or more can be identified,  a very interesting
constraint on $(\tilde{\kappa},\kappa)$ plane can be obtained by requiring
the cross section of $t\bar tj$ production to be less than $t\bar t$ 
production, i.e., by requiring the ratio to be less than 1.
More importantly, by requiring $\sigma(t\bar tj)$ with a $p_T(j)$ cut to be 
less than a certain fraction of $\sigma(t\bar t)$, of which both cross
sections can be measured in experiments, $(\tilde\kappa, \kappa)$ can be 
further constrained.

\subsection{Differential Cross Sections}

We shall next examine the effects of $\tilde\kappa$ and $\kappa$ on 
differential cross sections. We first show in Fig.~\ref{fig7} the spectra of 
the transverse momentum of the top quark for some typical values of 
$(\tilde \kappa, \kappa)=[(0,0),(1,0),(0,1),(0,-0.5),(1,1)]$.  Part (a) is for 
$t\bar t$ production, while (b), (c), and (d) are for $t\bar tj$ production
with $p_T(j)>5,10$, and 20 GeV, respectively. 
{}From Fig.~\ref{fig7} we observe the following common features for both
$t\bar t$ and $t\bar t j$ production.
The shape of the transverse momentum $p_T$ spectrum of the top quark are 
not sensitive to the CEDM form factor $\tilde \kappa$, which only affects 
the normalization, as indicated by comparing the $(\tilde\kappa,\kappa)=(0,0)$ 
and $(1,0)$ curves and by comparing $(0,1)$ and $(1,1)$ curves.  
It is understood that the simple $p_T$ distribution is CP-conserving and,
therefore, not sensitive to the CP-violating CEDM form factor $\tilde\kappa$.  
We have also verified that negative and positive $\tilde\kappa$'s with the 
same magnitude produce the same $p_T$ spectrum.  
However, the CMDM form factor $\kappa$ affects the shape of the $p_T$ spectrum
in a non-trivial way.  A positive $\kappa$ enhances the $p_T$ spectrum
in the large $p_T$ region, thus, making the $p_T$ spectrum 
significantly harder than the SM result that should be detectable, 
as shown by the $(\tilde\kappa,\kappa)=(0,1),(1,1)$ curves 
in Fig.~\ref{fig7}.  
On the other hand, a negative $\kappa$ does not affect the shape of the $p_T$
spectrum appreciably, as shown by the $(0,-0.5)$ curve.  
Therefore, by measuring the $p_T$ spectrum information
on $\kappa$ can be obtained.
We also note that the relative shapes of the $p_T$ spectra are the same for
$t\bar t$ and $t\bar tj$ production with different $p_T(j)$ cuts.  In other
words, the effects of CEDM and CMDM form factors on the $p_T$ spectrum of 
$t\bar t$ production are about the same whether or not an extra jet is tagged.
In Fig.~\ref{fig8} we show the rapidity distribution of the top quark
for the same set of $(\tilde\kappa, \kappa)$.
The shape of the rapidity spectrum of the top quark is not sensitive to 
both $\tilde\kappa$ and $\kappa$ that only the normalization is affected, as
indicated in Fig.~\ref{fig8}.  Also, the relative positions of the $y$ 
spectra are about the same for $t\bar t$ and $t\bar tj$ production 
with various $p_T(j)$ cuts.
In Fig.~\ref{fig9} we show the transverse momentum and rapidity distributions
of the jet in the $t\bar tj$ production, with the same set of 
$(\tilde\kappa,\kappa)$.  Qualitatively, the various $(\tilde\kappa,\kappa)$ 
curves of the $p_T(j)$ and $y(j)$ distributions are similar in shape.
This fact explains why the relative shapes of the $p_T$ spectra
in Fig.~\ref{fig7} and the relative positions of the $y$ spectra in 
Fig.~\ref{fig8} are about the same for $t\bar t$ production and $t\bar tj$ 
production with different $p_T(j)$ cuts.

The conclusion of studying the total cross sections and 
differential distributions is that we did not gain better sensitivities to 
$\tilde\kappa$ and $\kappa$ by tagging an extra jet in $t\bar t$ production.
However, by requiring the cross section of $t\bar tj$ production to be less
than a certain fraction of $t\bar t$ production under a $p_T(j)$ cut, 
of which both cross sections can be measured in experiments,  one can 
constrain $(\tilde\kappa,\kappa)$.

\section{CP-Observables}

In the last section, we have shown that the distributions for CP-even
observables such as $p_T$ are not sensitive to $\tilde \kappa$.  
Moreover, higher order corrections can render the detection
by such distributions useless.  Only unless $\tilde \kappa$ is very large 
can the effects be detected.  Thus, in this section we shall look at some
CP-odd observables, which should be sensitive to the CEDM of the top quark
and safe from higher order corrections.
A nonzero expectation value for such a CP-odd observable at the Tevatron
should be a signal for CP-violation, because the initial state $p\bar p$ is 
a CP eigenstate and, also, its expectation value is not affected by 
the CP-even higher order corrections.

In our effective Lagrangian, since we have assumed $\tilde \kappa$ to be 
real only those CP-odd and $\hat T$-odd variables can probe $\tilde \kappa$.
CP-odd and $\hat T$-even variables can only probe the imaginary part of 
$\tilde \kappa$, but such an imaginary part must vanish at zero momentum
transfer, so it must be related to terms of higher dimension ($>5$) in
the effective Lagrangian.

In the last section, the cross sections for $t\bar t$ and $t\bar tj$ 
production are shown for the case that the top helicities are summed.  From 
Eqs.~(\ref{tt1})--(\ref{tt2}) or from the contour plots we can see that
the total cross sections do not contain terms linearly proportional to
$\tilde \kappa$.  Therefore, the total cross section is not sensitive to
CP-odd observables when the top helicities are summed.  However,
it was shown in Ref.~\cite{aas} that when the top helicities are not summed
the cross section does have terms linearly proportional to $\tilde \kappa$.
In other words, in order to detect the CP-violating effects due to $\tilde
\kappa$ one needs to have information about the top helicities.  
Fortunately, the top is so heavy that it decays before hadronization takes
place \cite{bigi} and, therefore, the spin information of the top quark
is retained in the decay products.   The top
helicities or polarizations are not directly measured but can be realized
in its weak decay \cite{darwin,schmidt}, 
because of the left-handed nature of the weak interaction.
Since the top is heavy, in its rest frame the top quark first decays into
a $b$ quark and $W^+$ boson, with the $b$ preferentially left-handed and the 
$W^+$ boson predominately longitudinal.  Due to angular momentum conservation
the longitudinal $W^+$ boson is preferentially produced along with the 
direction of the top quark polarization.  Therefore, the anti-lepton
$\ell^+$ produced in the $W^+$ decay also prefers to be in the direction
of the top polarization.   Similarly, the momentum of the lepton $\ell^-$ 
produced in the anti-top decay prefers to be in the opposite direction of the 
anti-top polarization.    Thus, by discriminating the directions of the 
lepton and anti-lepton one can select particular polarizations of the top 
and anti-top.  Similarly, this argument can also be applied to the $b$ quark
and $\bar b$ antiquark.   We shall look at the following variables 
\cite{aas,ma}:
\begin{eqnarray}
v_1 &=& {\bf \hat p} \cdot ( {\bf l^+} \times {\bf l^-} )\;
        {\bf \hat p} \cdot ( {\bf l^+} - {\bf l^-} ) /m_t^3\;, \\
v_2 &=& ( {\bf l^+} - {\bf l^-} ) \cdot ( {\bf b} \times {\bf \bar b})
        /m_t^3 \;, \\
v_3 &=& \epsilon_{\mu\nu\rho\sigma}\, {\ell^+}^\mu \, {\ell^-}^{\nu}\,
        b^\rho\, \bar b^\sigma /m_t^4 \;, \\
v_4 &=& {\bf \hat p} \cdot ( {\bf b} \times {\bf \bar b} )\;
        {\bf \hat p} \cdot ( {\bf b} - {\bf \bar b} ) /m_t^3\;,
\end{eqnarray}
where $\epsilon_{0123}=-1$, and ${\bf l^+, l^-, b, \bar b}$ represent
the 3-momenta of the leptons and quarks , and $\ell^+, \ell^-,b,\bar b$ 
represent the 4-momenta of the leptons and quarks, and ${\bf \hat p}$ is
the unit vector of the proton.
Since the initial state of the collision is $p\bar p$, which is a 
CP-eigenstate, nonzero expectation values for these variables are signals
of CP-violation.  Some of these variables, as signals of CP-violation due to
the CEDM of the top, have been demonstrated for $t\bar t$ production in
Ref.~\cite{aas,ma}.  Our aim is to compare the sensitivities of the 
$t\bar t$ and $t\bar tj$ production to these CP-odd observables.
We shall put $\kappa=0$ in the following, as it
does not affect the expectation value of these CP-odd observables.

We put in the semileptonic decays of the top and anti-top using the helicity
amplitude method with full spin correlation (described in the appendix.)
In order to detect the leptons and quarks we impose a set of minimal
cuts:
\begin{equation}
p_T(\ell,b)>5\;{\rm GeV} \qquad |y(\ell,b)|<2 \;.
\end{equation}
We calculate numerically the expectation values of these observables 
for $\tilde \kappa$ between $-1$ and 1 with an increment of 0.1 or 0.2.
The asymmetry $A_v$ for a variable $v$ is defined as 
\begin{equation}
A_v = \frac{\langle v \rangle }{\sqrt{ \langle v^2 \rangle}}
\end{equation}
where the expectation value $\langle v \rangle =\int v \frac{d\sigma}{dv} dv/
\int \frac{d\sigma}{dv} dv$.  Therefore, the expectation values $\langle v 
\rangle$ and $\langle v^2 \rangle$ do not scale with the total cross sections
or with the branching ratios of the top and anti-top quarks.  \
The number of signal events due to this asymmetry is $NA_v$, where $N$ is the
total number of events, and $\sqrt{N}$ is the standard deviation of 
statistical fluctuation.  Therefore, the condition for the signal of 
the asymmetry to have a $\eta$ sigma significance is given by
\begin{equation}
\label{**}
\sqrt{N} \, A_v \ge \eta \qquad {\rm or} \qquad \sqrt{N} \, 
\frac{\langle v \rangle}{\sqrt{\langle v^2 \rangle}} \ge \eta  \;.
\end{equation}

We shall first give the results for $t\bar t$ production.  For $\tilde \kappa$
between $-1$ and 1, the expectation values $\langle v_i \rangle$ and 
$\sqrt{\langle v_i^2 \rangle}$ can be expressed as
\begin{equation}
\label{*}
\begin{array}{rcl}
\langle v_1 \rangle = -0.0037 \tilde{\kappa} +7.3\times 10^{-5} 
\tilde{\kappa}^2 + 0.0015 \tilde{\kappa}^3\,, && 
\sqrt{\langle v_1^2 \rangle} \simeq  0.046 \;,  \\
\langle v_2 \rangle = 0.0040 \tilde{\kappa} - 1.7\times 10^{-5} 
\tilde{\kappa}^2 - 0.0015 \tilde{\kappa}^3\,, && 
\sqrt{\langle v_2^2 \rangle} \simeq  0.11 \;,  \\
\langle v_3 \rangle = 0.0056 \tilde{\kappa} -7.6\times 10^{-5} 
\tilde{\kappa}^2 - 0.0021 \tilde{\kappa}^3\,, && 
\sqrt{\langle v_3^2 \rangle} \simeq  0.07 \;,  \\
\langle v_4 \rangle = -0.0018 \tilde{\kappa} + 7.4\times 10^{-5} 
\tilde{\kappa}^2 + 0.00075 \tilde{\kappa}^3\,, && 
\sqrt{\langle v_4^2 \rangle} \simeq  0.09 \; ,
\end{array}
\end{equation}
where we fitted the numerical results using a polynomial in $\tilde \kappa$
up to $\tilde {\kappa}^3$.  For sufficiently small $\tilde\kappa$ only the
first term linear in $\tilde \kappa$ is important.  The $\sqrt{\langle
v_i^2 \rangle}$  are roughly independent of $\tilde\kappa$ for 
$\tilde\kappa$ between $-1$ and 1.
To test the sensitivity of each observable to $\tilde\kappa$ we require a
1 sigma effect ($\eta=1$) of the signal, and by Eq.~(\ref{**}) 
$|\tilde\kappa|$ must be larger than a minimum value, given by 
\begin{equation}
\label{18}
|\tilde \kappa| \ge \left\{ \begin{array}{ll}
      \frac{12}{\sqrt N} & {\rm for}\; v_1 \\
      \frac{27}{\sqrt N} & {\rm for}\; v_2 \\
      \frac{12}{\sqrt N} & {\rm for}\; v_3 \\
      \frac{50}{\sqrt N} & {\rm for}\; v_4
                           \end{array}
\right.  \;,
\end{equation}
where we assumed only the linear term in $\tilde\kappa$ in Eq.~(\ref{*}).
We can immediately see that the observables $v_1$ and $v_3$ are about the 
same in sensitivity to $\tilde\kappa$, and more sensitive than $v_2$ and 
$v_4$.  

For $t\bar tj$ production we only give results on $v_1$ and $v_3$ 
with $p_T(j)>5,10,20$ GeV.  The results for the expectation values
of $v_1$ and $v_3$ are:
\begin{equation}
\langle v_1 \rangle = \left\{ \begin{array}{ll}
     - 0.0032\tilde\kappa - 1.4\times 10^{-4} \tilde\kappa^2 +0.0020 \tilde
\kappa^3\; & {\rm for}\; p_T(j)>5\,{\rm GeV} \\
     - 0.0028\tilde\kappa - 2.1\times 10^{-4} \tilde\kappa^2 +0.0016 \tilde
\kappa^3\; & {\rm for}\; p_T(j)>10\,{\rm GeV} \\
     - 0.0030\tilde\kappa - 7.7\times 10^{-6} \tilde\kappa^2 +0.0019 \tilde
\kappa^3\; & {\rm for}\; p_T(j)>20\,{\rm GeV} 
                         \end{array} \right. ,
\qquad \sqrt{ \langle v_1^2 \rangle } \simeq 0.046
\end{equation}
\begin{equation}
\langle v_3 \rangle = \left\{ \begin{array}{ll}
      0.0032\tilde\kappa - 3.6\times 10^{-4} \tilde\kappa^2 -0.0015 \tilde
\kappa^3\; & {\rm for}\; p_T(j)>5\,{\rm GeV} \\
      0.0033\tilde\kappa - 3.7\times 10^{-4} \tilde\kappa^2 -0.0018 \tilde
\kappa^3\; & {\rm for}\; p_T(j)>10\,{\rm GeV} \\
      0.0032\tilde\kappa - 3.6\times 10^{-4} \tilde\kappa^2 -0.0018 \tilde
\kappa^3\; & {\rm for}\; p_T(j)>20\,{\rm GeV} 
                         \end{array} \right. 
\qquad \sqrt{ \langle v_3^2 \rangle } \simeq 0.07
\end{equation}
We can see that the sensitivities of $\langle v_1 \rangle$ and 
$\langle v_3 \rangle$ are about the same for various $p_T(j)$ cuts
at small $\tilde\kappa$ .   But, both $\langle v_1 \rangle$ and 
$\langle v_3 \rangle$  are less sensitive to $\tilde\kappa$ 
in $t\bar tj$ production than in $t\bar t$ production, since the 
expectation value
$\langle v \rangle$ is getting smaller while $\sqrt{\langle v^2 \rangle }$
remains the same.    We have also verified that $v_2$
and $v_4$ have similar behavior.  
In general, the sensitivities of the CP-odd observables under consideration
decrease when going from $t\bar t$ production to $t\bar tj$ production,
especially, for the observables $v_2$, $v_3$, and $v_4$ that require the 
$b$ quark and $\bar b$ antiquark momenta.

\section{Conclusions}

In this paper, we have studied the effects of the CEDM and CMDM couplings of
the top quark on the $t\bar t$ and $t\bar tj$ production, as well as 
the ratio of these two cross sections at the Tevatron with $\sqrt{s}=2$ TeV.
We found that by demanding $\sigma(t\bar tj)$ to be less than a certain
fraction of $\sigma(t\bar t)$ we can obtain constraints on $\tilde\kappa$ and 
$\kappa$.  
We have also 
shown that the shape of the differential distributions ($p_T$ and $y$)
is not sensitive to the CEDM form factor $\tilde\kappa$ because these
distributions are CP-even. 
However, the $p_T$ distribution is very sensitive to 
the sign of the CMDM form factor $\kappa$.  A positive $\kappa$ 
significantly hardens the $p_T$ spectrum while a negative $\kappa$ does not.
Therefore, by measuring the $p_T$ spectrum of the top quark we can put
bounds on $\kappa$.

Furthermore, we have also studied the effects of $\tilde\kappa$ on the
expectation values of some CP-odd and $\hat T$-odd observables
in both $t\bar t$ and $t\bar tj$ production.
The asymmetry $A_{v}$ obtained for these observables ranges between 
$(2-8)\times 10^{-2} \tilde\kappa$.   The SM cross section for $t\bar t$ 
production with both the top and anti-top decaying semileptonically is
about 0.25 pb.  With a luminosity of, say, 5 fb$^{-1}$, there are totally
of order 1200 events.  Using 1-sigma effect as the discovery 
criterion, it can probe
the region $|\tilde\kappa| \agt 0.35$ (using Eq.~(\ref{18})).  In other 
words, if no effect is observed, we can constrain $|\tilde\kappa| \alt 0.35$.
Also, we found that the sensitivities obtained in
$t\bar tj$ production are smaller than in $t\bar t$ production.  Therefore,
it is not advantageous to tag an extra jet in $t\bar t$ production with 
respect to these CP-odd observables.  
Since it is very often to have extra jets in $t\bar t$ production, the 
results obtained for the CP-odd observables in $t\bar t$ production will be
very often contaminated by the smaller results of the $t\bar tj$ production.
In reality, the experimental measurement will be somewhere in between 
the results of $t\bar t$ and $t\bar tj$ production.
In this work, we do not consider the effects of gluon radiating off the 
decay products of the top quark.  These effects will also be complicated 
by the finite width of the top quark.  However, we do not expect any
significant changes on our conclusions.

\bigskip

\section*{Acknowledgement}

I thank Duane Dicus, Roberto Vega, Tzu Chiang Yuan, and David Bowser-Chao 
for useful discussions. 
This work was supported by the U.~S. Department of Energy, Division of
High Energy Physics, under Grant DE-FG03-93ER40757.
\newpage
\appendix
\section{}

In this appendix we shall list all helicity amplitudes for the processes
(i) $q\bar q \to Q \bar Q g$, and (ii) $gg \to Q \bar Q g$, where $Q$ stands
for $t$ or $b$.

\subsection{$q_i(p_1)\; \bar q_j (p_2)\; \to Q_k (k_1) \bar Q_l (k_2) g_a (g)$}

The momenta of the particles are labeled in the parentheses and 
the subscripts denote the color indices of the quarks and the gluon.  The
contributing Feynman diagrams are shown in Fig.~\ref{fig2}.  We list
the helicity amplitudes for each diagram but without the color factors, which
we shall sum later. We use the following short-hand notations:
\begin{equation}
\hat s = (p_1+p_2)^2\,, \qquad \hat s' = (k_1+k_2)^2\,, \qquad J_\rho =
\bar v(p_2) \gamma_\rho u(p_1) \;,
\end{equation}
\begin{equation}
\Gamma_\mu(p,q; \epsilon_1, \epsilon_2) = (p-q)_\mu\, 
\epsilon_1 \cdot \epsilon_2 + (p+2q) \cdot \epsilon_1 \;\epsilon_{2\mu} -
(2p+q)\cdot \epsilon_2 \;\epsilon_{1\mu}  \;.
\end{equation}
We also define the following
\begin{equation}
S(p,q) = \overlay{/}{p} +\frac{1}{4m_Q} \left( \overlay{/}{p} \overlay{/}{q} -
\overlay{/}{q} \overlay{/}{p} \right ) \left( \kappa - i \tilde{\kappa} 
\gamma^5 \right )
\end{equation}
where $p,q$ are 4-vectors, or if $p$ or $q$ is an index it represents a 
gamma matrix.
The helicity amplitudes are given by
\begin{eqnarray}
{\cal M}_a &=& \frac{g_s^3}{\hat s} \bar u(k_1)\, S(\epsilon(g), -g)\, \frac{
\overlay{/}{k}_1+\overlay{/}{g} + m_Q}{(k_1+g)^2 - m_Q^2}\,S(\rho, p_1+p_2)\,
v(k_2)\; J_\rho \;, \\
{\cal M}_b &=& \frac{g_s^3}{\hat s} \bar u(k_1)\, S(\rho, p_1+p_2)\, \frac{
-\overlay{/}{k}_2-\overlay{/}{g} + m_Q}{(k_2+g)^2 - m_Q^2}\,
S(\epsilon(g), -g)\, v(k_2)\; J_\rho \;, \\
{\cal M}_c &=& \frac{g_s^3}{\hat s \hat s'} \bar u(k_1)\, S(\rho, k_1+k_2)\, 
v(k_2)\; \Gamma_\rho \left( p_1+p_2, -g; J, \epsilon(g) \right)\;, \\
{\cal M}_d &=& \frac{g_s^3}{\hat s'} \bar u(k_1)\, S(\rho, k_1+k_2)\, v(k_2)\;
\bar v(p_2) \gamma_\rho \, \frac{\overlay{/}{p}_1-\overlay{/}{g}}
{(p_1-g)^2}\, \overlay{/}{\epsilon}(g) \, u(p_1)\;, \\
{\cal M}_e &=& \frac{g_s^3}{\hat s'} \bar u(k_1)\, S(\rho, k_1+k_2)\, v(k_2)\;
\bar v(p_2)\, \overlay{/}{\epsilon}(g) \, \frac{-\overlay{/}{p}_2+
\overlay{/}{g}}{(-p_2+g)^2}\,\gamma_\rho \,  u(p_1)\;, \\
{\cal M}_f &=& \frac{g_s^3}{4m_Q \hat s} \bar u(k_1)\, 
\left( \overlay{/}{\epsilon}(g) \gamma^\rho - \gamma^\rho 
\overlay{/}{\epsilon}(g) \right ) \left(\kappa -i \tilde{\kappa} \gamma^5\right)
\, v(k_2) \; J_\rho \;,
\end{eqnarray}
where $\epsilon(g)$ is the polarization 4-vector of the gluon.
Taking into account the color factors the total amplitude can be written as
\begin{displaymath}
{\cal M} = \sum_{\alpha=1}^{4} O_\alpha {\cal M}_\alpha
\end{displaymath}
with 
\begin{eqnarray}
O_1 = (T^a T^b)_{kl} T^b_{ji}\,, &\qquad & 
O_3 = T^b_{kl} (T^bT^a)_{ji} \,, \nonumber\\
O_2 = (T^b T^a)_{kl} T^b_{ji}\,, &\qquad & 
O_4 = T^b_{kl} (T^aT^b)_{ji} \,,\nonumber
\end{eqnarray}
and 
\begin{eqnarray}
{\cal M}_1 = {\cal M}_a + {\cal M}_f\,, &\qquad & {\cal M}_3 = {\cal M}_d - 
{\cal M}_c\,, \nonumber \\
{\cal M}_2 = {\cal M}_b - {\cal M}_f\,, &\qquad & {\cal M}_4 = {\cal M}_e +
{\cal M}_c\,. \nonumber 
\end{eqnarray}
After squaring and summing all the color factors we have
\begin{eqnarray}
\sum_{\rm color} |{\cal M}|^2 &=& \frac{8}{3} \left( |{\cal M}_1|^2 
+ |{\cal M}_2|^2 + |{\cal M}_3|^2 + |{\cal M}_4|^2  \right ) \nonumber \\
&+& \frac{7}{3} \left( {\cal M}_1 {\cal M}_3^* + {\cal M}_1^* {\cal M}_3 +
                     {\cal M}_2 {\cal M}_4^* + {\cal M}_2^* {\cal M}_4 \right)
\nonumber \\
&-& \frac{2}{3} \left( {\cal M}_1 {\cal M}_4^* + {\cal M}_1^* {\cal M}_4 +
                     {\cal M}_2 {\cal M}_3^* + {\cal M}_2^* {\cal M}_3 \right) 
\nonumber \\
&-& \frac{1}{3} \left( {\cal M}_1 {\cal M}_2^* + {\cal M}_1^* {\cal M}_2 +
                     {\cal M}_3 {\cal M}_4^* + {\cal M}_3^* {\cal M}_4 \right)
\end{eqnarray}

\subsection{$g_a(g_1)\, g_b(g_2) \, \to Q_i (k_1)\, \bar Q_j(k_2) \, g_c(g)$}

The contributing Feynman diagrams are shown in Fig.~\ref{fig3}.  The 
helicity amplitudes without the color factors for each diagram are given by
\begin{eqnarray}
{\cal M}_a &=& - g_s^3 \bar u(k_1)\, S(\epsilon(g), -g) \, 
\frac{\overlay{/}{k}_1 + \overlay{/}{g} + m_Q}{(k_1+g)^2 - m_Q^2}\,
S(\epsilon(g_1),g_1)\,\frac{-\overlay{/}{k}_2 + \overlay{/}{g}_2 + m_Q}
{(k_2-g_2)^2 - m_Q^2}\, S(\epsilon(g_2), g_2)\, v(k_2) \;, \\
{\cal M}_b &=& - g_s^3 \bar u(k_1)\, S(\epsilon(g_1), g_1) \, 
\frac{\overlay{/}{k}_1 - \overlay{/}{g}_1 + m_Q}{(k_1-g_1)^2 - m_Q^2}\,
S(\epsilon(g),-g)\,\frac{-\overlay{/}{k}_2 + \overlay{/}{g}_2 + m_Q}
{(k_2-g_2)^2 - m_Q^2}\, S(\epsilon(g_2), g_2)\, v(k_2) \;, \\
{\cal M}_c &=& - g_s^3 \bar u(k_1)\, S(\epsilon(g_1), g_1) \, 
\frac{\overlay{/}{k}_1 - \overlay{/}{g}_1 + m_Q}{(k_1-g_1)^2 - m_Q^2}\,
S(\epsilon(g_2),g_2)\,\frac{-\overlay{/}{k}_2 - \overlay{/}{g} + m_Q}
{(k_2+g)^2 - m_Q^2}\, S(\epsilon(g), -g)\, v(k_2) \;, \\
{\cal M}_d &=& - \frac{g_s^3}{(g_1-g)^2} \, \bar u(k_1)\, S(\rho, g_1-g) \, 
\frac{-\overlay{/}{k}_2 + \overlay{/}{g}_2 + m_Q}{(k_2-g_2)^2 - m_Q^2}\,
S(\epsilon(g_2), g_2)\, v(k_2) \; \Gamma_\rho( g_1, -g; \epsilon(g_1), 
\epsilon(g) ) \;, \\
{\cal M}_e &=& - \frac{g_s^3}{(g_2-g)^2} \, \bar u(k_1)\, S(\epsilon(g_1), g_1)\, \frac{\overlay{/}{k}_1 - \overlay{/}{g}_1 + m_Q}{(k_1-g_1)^2 - m_Q^2}\,
S(\rho, g_2-g)\, v(k_2) \; \Gamma_\rho( g_2, -g; \epsilon(g_2),\epsilon(g)) \;,
\\
{\cal M}_f &=& - \frac{g_s^3}{4m_Q} \, \bar u(k_1)\, \left( 
\overlay{/}{\epsilon}(g) \overlay{/}{\epsilon}(g_1) -
\overlay{/}{\epsilon}(g_1) \overlay{/}{\epsilon}(g) \right) \left(\kappa -
i \tilde{\kappa} \gamma^5 \right) \, 
\frac{-\overlay{/}{k}_2 + \overlay{/}{g}_2 + m_Q}{(k_2-g_2)^2 - m_Q^2}\,
S(\epsilon(g_2), g_2)\, v(k_2) \; \\
{\cal M}_g &=& - \frac{g_s^3}{4m_Q} \, \bar u(k_1)\, S(\epsilon(g_1), g_1)\,
\frac{\overlay{/}{k}_1 - \overlay{/}{g}_1 + m_Q}{(k_1-g_1)^2 - m_Q^2}\,
\left( 
\overlay{/}{\epsilon}(g_2) \overlay{/}{\epsilon}(g) -
\overlay{/}{\epsilon}(g) \overlay{/}{\epsilon}(g_2) \right) \left(\kappa -
i \tilde{\kappa} \gamma^5 \right) \,  v(k_2) \; \\
{\cal M}_h &=& - \frac{g_s^3}{(g_1+g_2)^2} \, \bar u(k_1)\, S(\epsilon(g),-g)
\, \frac{\overlay{/}{k}_1 + \overlay{/}{g} + m_Q}{(k_1+g)^2 - m_Q^2}\,
S(\rho, g_1+g_2)\, v(k_2) \; \Gamma_\rho(g_1,g_2; 
\epsilon(g_1), \epsilon(g_2) ) \;, \\
{\cal M}_i &=& - \frac{g_s^3}{(g_1+g_2)^2} \, \bar u(k_1)\, S(\rho, g_1+g_2)
\, \frac{-\overlay{/}{k}_2 - \overlay{/}{g} + m_Q}{(k_2+g)^2 - m_Q^2}\,
S(\epsilon(g), -g)\, v(k_2) \; \Gamma_\rho(g_1,g_2; \epsilon(g_1), 
\epsilon(g_2) ) \;, \\
{\cal M}_j &=& - \frac{g_s^3}{4m_Q (g_1+g_2)^2} \, \bar u(k_1)\, \left(
\gamma^\rho \overlay{/}{\epsilon}(g) - \overlay{/}{\epsilon}(g) \gamma^\rho 
\right) \left( \kappa - i \tilde{\kappa}\gamma^5 \right )\,
v(k_2) \; \Gamma_\rho(g_1,g_2; \epsilon(g_1), \epsilon(g_2) )\;, \\
{\cal M}_k &=& - \frac{g_s^3}{(k_1+k_2)^2 (g_1+g_2)^2} \, \bar u(k_1)\, 
S(\rho, k_1+k_2) \, v(k_2) \; \Gamma_\rho(-g, g_1+g_2; \epsilon(g), \Gamma_1 )
\;, \\
&& \qquad \qquad \qquad \qquad 
{\rm with} \; \Gamma_{1\mu}=\Gamma_\mu(g_1,g_2;\epsilon(g_1),
\epsilon(g_2) ) \nonumber \\
{\cal M}_l &=& - \frac{g_s^3}{(k_1+k_2)^2 (g_1-g)^2} \, \bar u(k_1)\, 
S(\rho, k_1+k_2) \, v(k_2) \; \Gamma_\rho(g_2, g_1-g; \epsilon(g_2), \Gamma_2 )
\;, \\
&& \qquad \qquad \qquad \qquad {\rm with} \; 
\Gamma_{2\mu}=\Gamma_\mu(g_1,-g;\epsilon(g_1), \epsilon(g) ) \nonumber \\
{\cal M}_m &=& - \frac{g_s^3}{(k_1+k_2)^2 (g_2-g)^2} \, \bar u(k_1)\, 
S(\rho, k_1+k_2) \, v(k_2) \; \Gamma_\rho(g_1, g_2-g; \epsilon(g_1), \Gamma_3 )
\;, \\
&& \qquad \qquad \qquad \qquad {\rm with} \; 
\Gamma_{3\mu}=\Gamma_\mu(g_2,-g;\epsilon(g_2),
\epsilon(g) ) \nonumber \\
{\cal M}_{n_1} &=& -\frac{g_s^3}{(k_1+k_2)^2}\, \bar u(k_1)\, S(\rho,k_1+k_2)
\, v(k_2)\; \epsilon(g_1)\cdot \epsilon(g)\, \epsilon(g_2)_\rho \;, \\
{\cal M}_{n_2} &=& -\frac{g_s^3}{(k_1+k_2)^2}\, \bar u(k_1)\, S(\rho,k_1+k_2)
\, v(k_2)\; \epsilon(g_2)\cdot \epsilon(g)\, \epsilon(g_1)_\rho \;, \\
{\cal M}_{n_3} &=& -\frac{g_s^3}{(k_1+k_2)^2}\, \bar u(k_1)\, S(\rho,k_1+k_2)
\, v(k_2)\; \epsilon(g_1)\cdot \epsilon(g_2)\, \epsilon(g)_\rho \;, \\
{\cal M}_o &=& -\frac{g_s^3}{4m_Q}\, \bar u(k_1)\, S(\epsilon(g),-g)\,
\frac{\overlay{/}{k}_1 + \overlay{/}{g} + m_Q}{(k_1+g)^2 - m_Q^2}\,
\left( \overlay{/}{\epsilon}(g_1) \overlay{/}{\epsilon}(g_2) - 
\overlay{/}{\epsilon}(g_2)\overlay{/}{\epsilon}(g_1) \right )
\left(\kappa-i\tilde{\kappa}\gamma^5 \right )\, v(k_2)\;  \\
{\cal M}_p &=& -\frac{g_s^3}{4m_Q}\, \bar u(k_1)\, 
\left( \overlay{/}{\epsilon}(g_1) \overlay{/}{\epsilon}(g_2) - 
\overlay{/}{\epsilon}(g_2)\overlay{/}{\epsilon}(g_1) \right )
\left(\kappa-i\tilde{\kappa}\gamma^5 \right )\, 
\frac{-\overlay{/}{k}_2 - \overlay{/}{g} + m_Q}{(k_2+g)^2 - m_Q^2}\,
S(\epsilon(g),-g)\, v(k_2)\; \\
{\cal M}_q &=& -\frac{g_s^3}{4m_Q(g_1-g)^2}\, \bar u(k_1)\, 
\left( \gamma^\rho \overlay{/}{\epsilon}(g_2) - \overlay{/}{\epsilon}(g_2) 
\gamma^\rho \right ) \left(\kappa-i\tilde{\kappa}\gamma^5 \right )\, v(k_2)\;
\Gamma_\rho( g_1, -g; \epsilon(g_1), \epsilon(g) )\;, \\
{\cal M}_r &=& -\frac{g_s^3}{4m_Q(g_2-g)^2}\, \bar u(k_1)\, 
\left( \overlay{/}{\epsilon}(g_1) \gamma^\rho  - \gamma^\rho 
\overlay{/}{\epsilon}(g_1)  \right ) \left(\kappa-i\tilde{\kappa}\gamma^5 
\right )\, v(k_2)\; \Gamma_\rho( g_2, -g; \epsilon(g_2), \epsilon(g) )\;,
\end{eqnarray}
where $\epsilon(g_1)$ and $\epsilon(g_2)$ are the polarization 4-vectors of
the gluons.
Also, we have to interchange the incoming gluons in diagrams (a)--(g) to make 
the complete set of Feynman diagrams:
\begin{equation}
{\cal M}'_{x=a,b,c,d,e,f,g} = {\cal M}_x \,(g_1 \leftrightarrow g_2 )\,.
\end{equation}
Since the diagram (n) has a quartic gluon vertex it is more convenient to
decompose it into 3 terms as we have ${\cal M}_{n_1},\,{\cal M}_{n_2},\,
{\cal M}_{n_3}$.   The complete amplitude including the color factors is given
by
\begin{displaymath}
{\cal M} = \sum_{\alpha=1}^6 O_\alpha {\cal M}_\alpha \;,
\end{displaymath}
where
\begin{eqnarray}
O_1 = (T^c T^a T^b )_{ij}\,, &\qquad & O_4 = (T^c T^b T^a)_{ij}\,, \nonumber \\
O_2 = (T^a T^c T^b )_{ij}\,, &\qquad & O_5 = (T^b T^c T^a)_{ij}\,, \nonumber \\
O_3 = (T^a T^b T^c )_{ij}\,, &\qquad & O_6 = (T^b T^a T^c)_{ij}\,, \nonumber 
\end{eqnarray}
and
\begin{eqnarray}
{\cal M}_1 &=& {\cal M}_a - {\cal M}_d + {\cal M}_f +{\cal M}_h -{\cal M}_j +
 {\cal M}_k +{\cal M}_l - {\cal M}_{n_1} + 2{\cal M}_{n_2} - {\cal M}_{n_3} +
{\cal M}_o - {\cal M}_q \nonumber \\
{\cal M}_2 &=& {\cal M}_b + {\cal M}_d - {\cal M}_e - {\cal M}_f -{\cal M}_g -
 {\cal M}_l -{\cal M}_m - {\cal M}_{n_1} - {\cal M}_{n_2} +2 {\cal M}_{n_3} +
{\cal M}_q - {\cal M}_r \nonumber \\
{\cal M}_3 &=& {\cal M}_c + {\cal M}_e + {\cal M}_g +{\cal M}_i +{\cal M}_j -
 {\cal M}_k +{\cal M}_m +2 {\cal M}_{n_1} -{\cal M}_{n_2} - {\cal M}_{n_3} +
{\cal M}_p + {\cal M}_r \nonumber \\
{\cal M}_4 &=& - {\cal M}_h + {\cal M}_j - {\cal M}_k +{\cal M}_m 
+2 {\cal M}_{n_1} -{\cal M}_{n_2} - {\cal M}_{n_3} - {\cal M}_o + {\cal M}_r +
{\cal M}_a' -{\cal M}_d' + {\cal M}_f' \nonumber \\
{\cal M}_5 &=& -{\cal M}_l - {\cal M}_m - {\cal M}_{n_1} -{\cal M}_{n_2} +2
 {\cal M}_{n_3} + {\cal M}_q - {\cal M}_r + {\cal M}_b' + {\cal M}_d' -
{\cal M}_e'- {\cal M}_f' -{\cal M}_g' \nonumber \\
{\cal M}_6 &=& -{\cal M}_i - {\cal M}_j + {\cal M}_k + {\cal M}_l - 
 {\cal M}_{n_1} +2 {\cal M}_{n_2} -  {\cal M}_{n_3} - {\cal M}_p - {\cal M}_q
 + {\cal M}_c' + {\cal M}_e' + {\cal M}_g' \nonumber \\
\end{eqnarray}
After squaring and summing the color factors, we have
\begin{eqnarray}
\sum_{\rm color} |{\cal M}|^2 &=& \frac{64}{9} \left( |{\cal M}_1|^2 +
|{\cal M}_2|^2 +|{\cal M}_3|^2 +|{\cal M}_4|^2 +|{\cal M}_5|^2 +|{\cal M}_6|^2
\right )  \nonumber \\
&-& \frac{8}{9} \biggr (  {\cal M}_1 {\cal M}_2^* + {\cal M}_1^* {\cal M}_2 +
     {\cal M}_3 {\cal M}_6^* + {\cal M}_3^* {\cal M}_6 +
{\cal M}_4 {\cal M}_5^* + {\cal M}_4^* {\cal M}_5 \nonumber \\ 
&& + {\cal M}_1 {\cal M}_4^* + {\cal M}_1^* {\cal M}_4 +
{\cal M}_2 {\cal M}_3^* + {\cal M}_2^* {\cal M}_3 +
{\cal M}_5 {\cal M}_6^* + {\cal M}_5^* {\cal M}_6  \biggr ) \nonumber\\
&+& \frac{10}{9} \biggr ( {\cal M}_1 {\cal M}_6^* + {\cal M}_1^* {\cal M}_6 +
{\cal M}_3 {\cal M}_4^* + {\cal M}_3^* {\cal M}_4 +
{\cal M}_2 {\cal M}_5^* + {\cal M}_2^* {\cal M}_5 \biggr ) \nonumber\\
&+& \frac{1}{9} \biggr ( {\cal M}_1 {\cal M}_3^* + {\cal M}_1^* {\cal M}_3 +
{\cal M}_2 {\cal M}_4^* + {\cal M}_2^* {\cal M}_4 +
{\cal M}_3 {\cal M}_5^* + {\cal M}_3^* {\cal M}_5 \nonumber \\
&& + {\cal M}_1 {\cal M}_5^* + {\cal M}_1^* {\cal M}_5 +
{\cal M}_2 {\cal M}_6^* + {\cal M}_2^* {\cal M}_6 +
{\cal M}_4 {\cal M}_6^* + {\cal M}_4^* {\cal M}_6  \biggr )
\end{eqnarray}

\subsection{$t \to b \ell^+ \nu$ and $\bar t\to \bar b \ell^- \bar \nu$}

These  decays can be included with full spin correlation using the helicity
amplitude method, by replacing 
\begin{eqnarray}
\bar u(t) &\to & - \frac{g^2}{8} \, \frac{1}{t^2 - m_t^2 + i \Gamma_t m_t} \,
 \frac{1}{W^2 -m_W^2 + i \Gamma_W m_W}\; \bar u(b)\, \overlay{/}{J}\,
(1-\gamma^5) \, (\overlay{/}{t} +m_t ) \;, \\
v(\bar t) & \to & - 
\frac{g^2}{8} \, \frac{1}{\bar t^2 - m_t^2 + i \Gamma_t m_t} \,
 \frac{1}{W^2 -m_W^2 + i \Gamma_W m_W}\; (- \overlay{/}{\bar t} +m_t ) \,
\overlay{/}{J'}\, (1-\gamma^5)\, v(\bar b) \;,
\end{eqnarray}
where
\begin{eqnarray}
J_\mu &=& \bar u(\nu) \,\gamma_\mu \, (1-\gamma^5) \, v(\ell^+) \,\nonumber \\
J'_\mu &=& \bar u(\ell^-) \,\gamma_\mu \, (1-\gamma^5) \, v(\bar \nu)
\nonumber \;,
\end{eqnarray}
and the momentum of each particle is labeled by the particle itself.
We use the narrow width approximation for the top and $W$ propagators.


\newpage
\begin{center}
{\bf \Large Figure Captions}
\end{center}

\begin{enumerate}
\item \label{fig1} Feynman diagrams for the process (a) $q\bar q\to Q \bar Q$,
and (b) $gg\to Q\bar Q$ with $Q=t$. 

\item \label{fig2} Contributing Feynman diagrams for the process $q\bar q \to
Q\bar Q g$ ($Q=t$).

\item \label{fig3} Contributing Feynman diagrams for the process $gg\to Q\bar
Q g$ ($Q=t$).  Diagrams (a)--(g) with the incoming gluons interchanged 
have to be included to make the complete set of diagrams.  

\item \label{fig4} Contours of the $t\bar t$ cross sections in pb on the 
$(\tilde{\kappa},\kappa)$ plane.

\item \label{fig5} Contours of the $t\bar t j$ cross sections in pb on the 
$(\tilde{\kappa},\kappa)$ plane, with $p_T(j) > 5, 10, 20$ GeV in (a), (b),
and (c), respectively.

\item \label{fig6} Contours of the ratio 
$\sigma(t\bar tj)/\sigma(t\bar t)$  on the 
$(\tilde{\kappa},\kappa)$ plane, with $p_T(j) > 5, 10, 20 $ GeV in (a), (b),
and (c), respectively.  Though the contours are rough, especially, in part (a)
due to some technical difficulties, the shape of the contours are rather 
clear.

\item \label{fig7} Differential cross section $d\sigma/dp_T$ versus the
transverse momentum of the top quark in (a) $t\bar t$ production, and in
$t\bar tj$ production with (b) $p_T(j)>5$ GeV, (c) $p_T(j)>10$ GeV, and (d)
$p_T(j)>20$ GeV.  
We show $(\tilde\kappa,\kappa)=(0,0)$ in solid, $(1,0)$ in dashes, 
$(0,1)$ in dots, $(0,-0.5)$ in dash-dot, and $(1,1)$ in dash-dot-dot.

\item \label{fig8} Differential cross section $d\sigma/dy$ versus the
rapidity of the top quark in (a) $t\bar t$ production, and in
$t\bar tj$ production with (b) $p_T(j)>5$ GeV, (c) $p_T(j)>10$ GeV, and (d)
$p_T(j)>20$ GeV.  
We show $(\tilde\kappa,\kappa)=(0,0)$ in solid, $(1,0)$ in dashes, 
$(0,1)$ in dots, $(0,-0.5)$ in dash-dot, and $(1,1)$ in dash-dot-dot.

\item \label{fig9} (a) Differential cross section $d\sigma/dp_T$ versus the
transverse momentum of the jet, and (b) differential cross section 
$d\sigma/dy$ versus the rapidity of the jet in 
$t\bar tj$ production.
\end{enumerate}

\pagestyle{empty}

\begin{figure}
\centering
\leavevmode
\epsfysize=475pt
\epsfbox{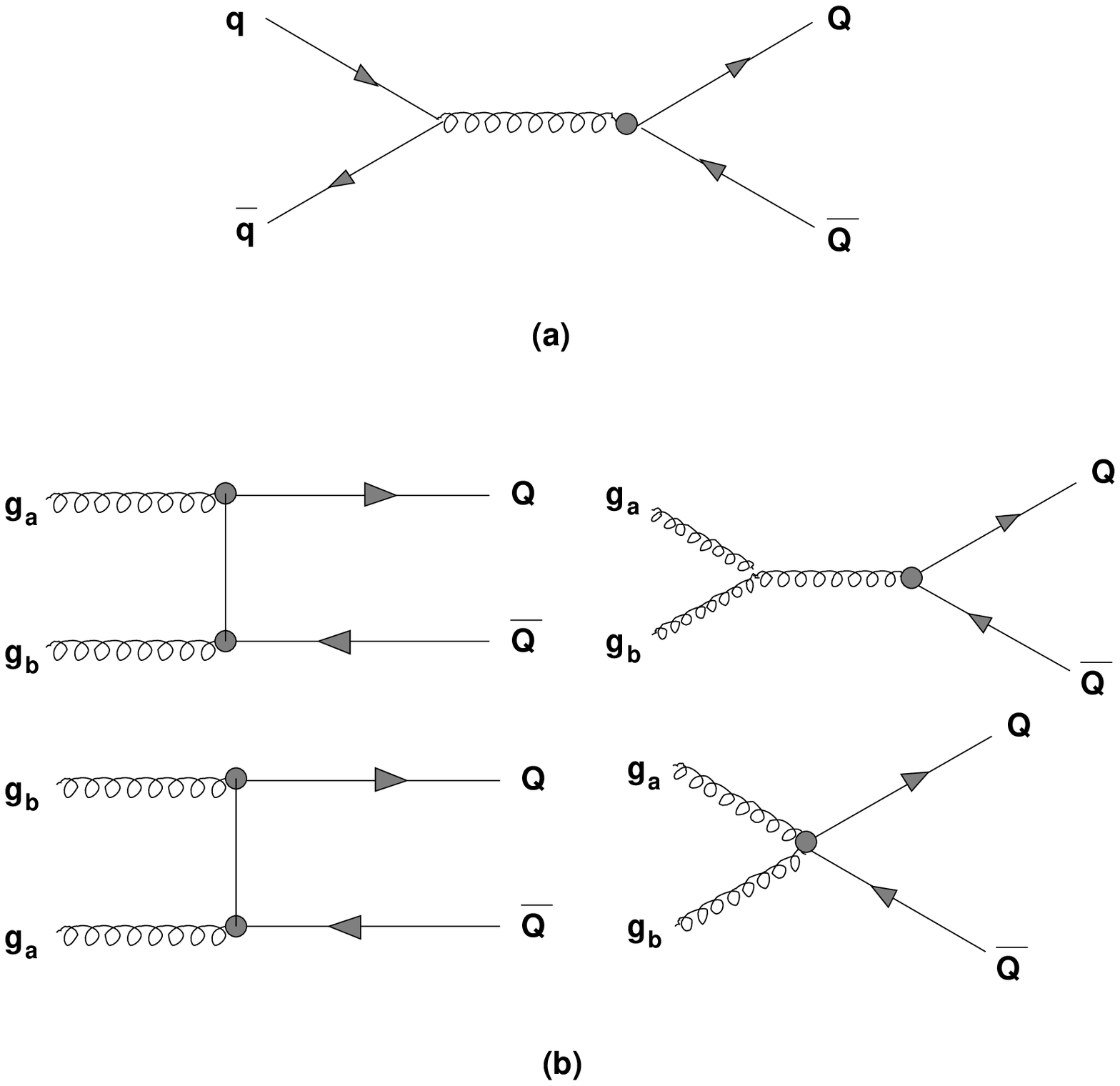}
\caption{}
\end{figure}

\begin{figure}
\centering
\leavevmode
\epsfysize=475pt
\epsfbox{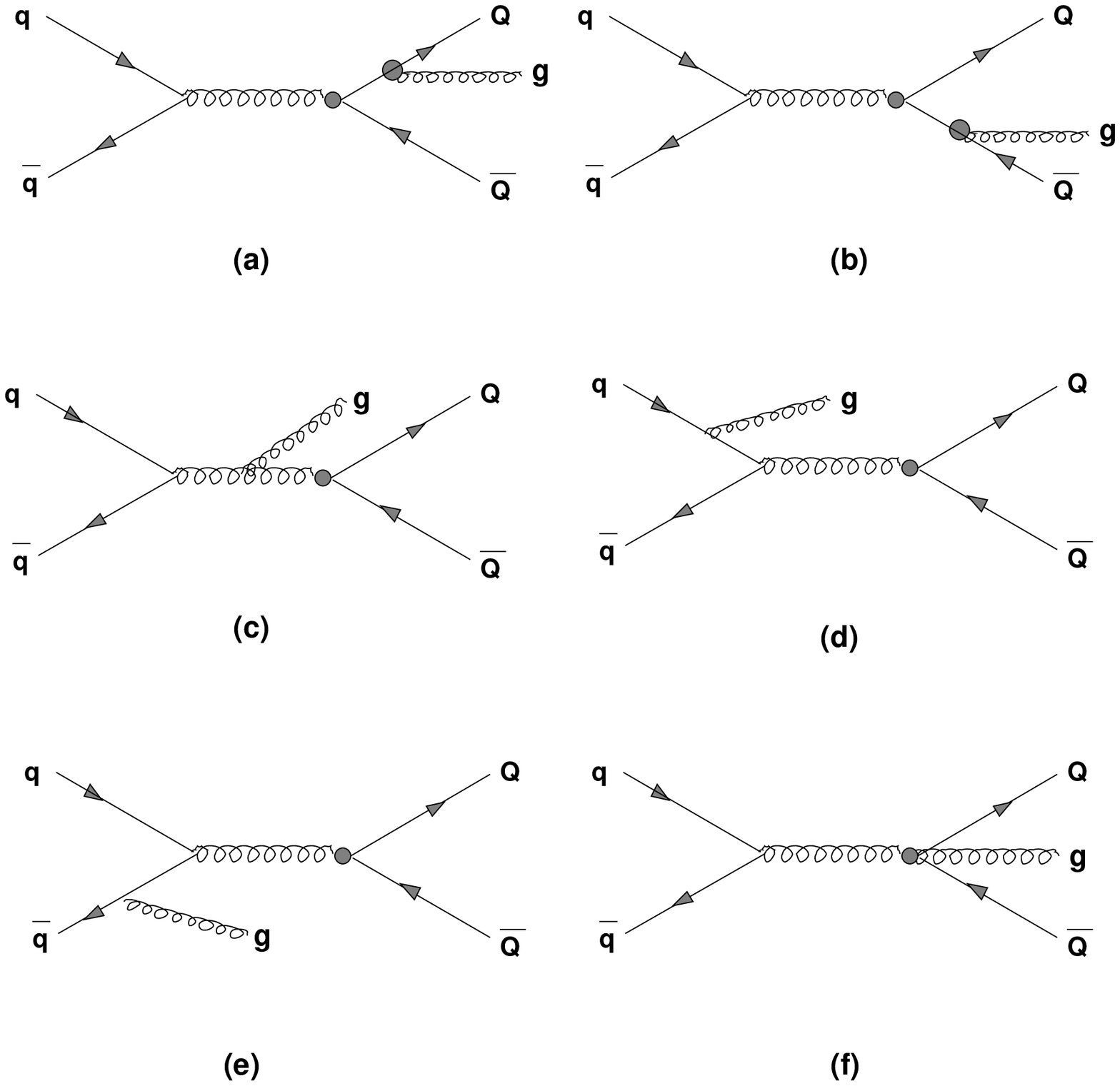}
\caption{}
\end{figure}

\begin{figure}
\centering
\leavevmode
\epsfysize=630pt
\epsfbox{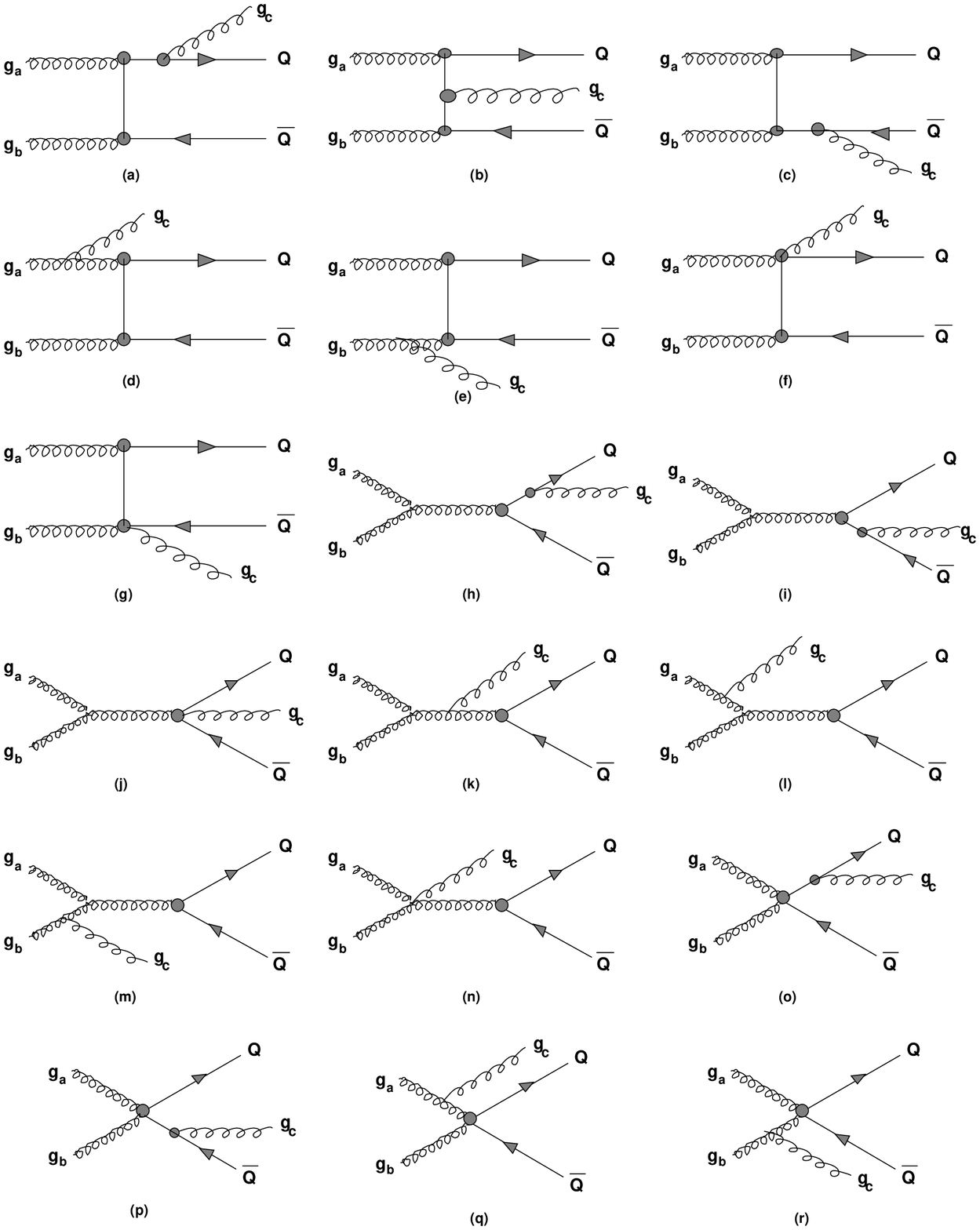}
\caption{}
\end{figure}

\begin{figure}
\centering
\leavevmode
\epsfysize=350pt
\epsfbox{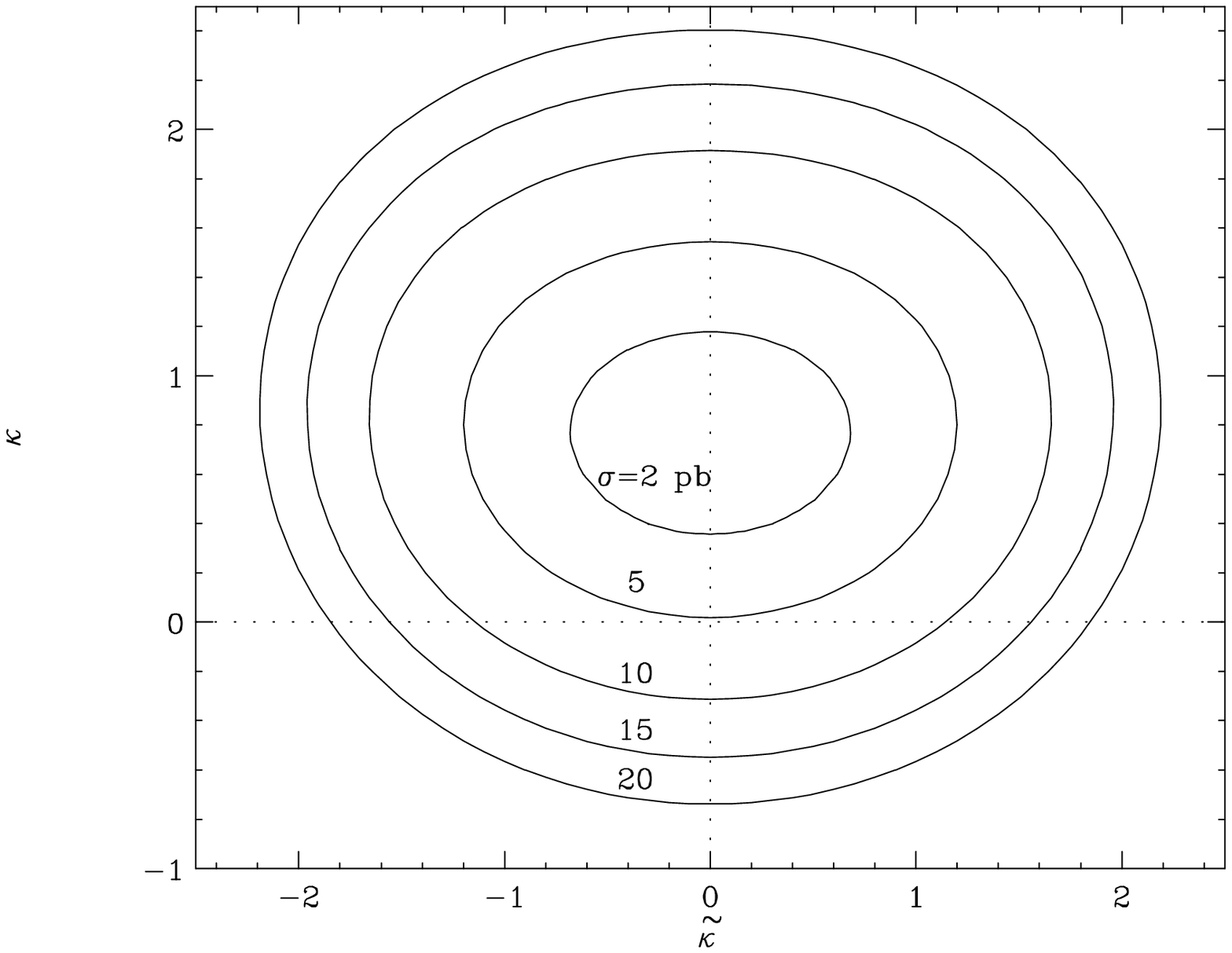}
\caption{}
\end{figure}

\begin{figure}
\centering
\leavevmode
\epsfysize=630pt
\epsfbox{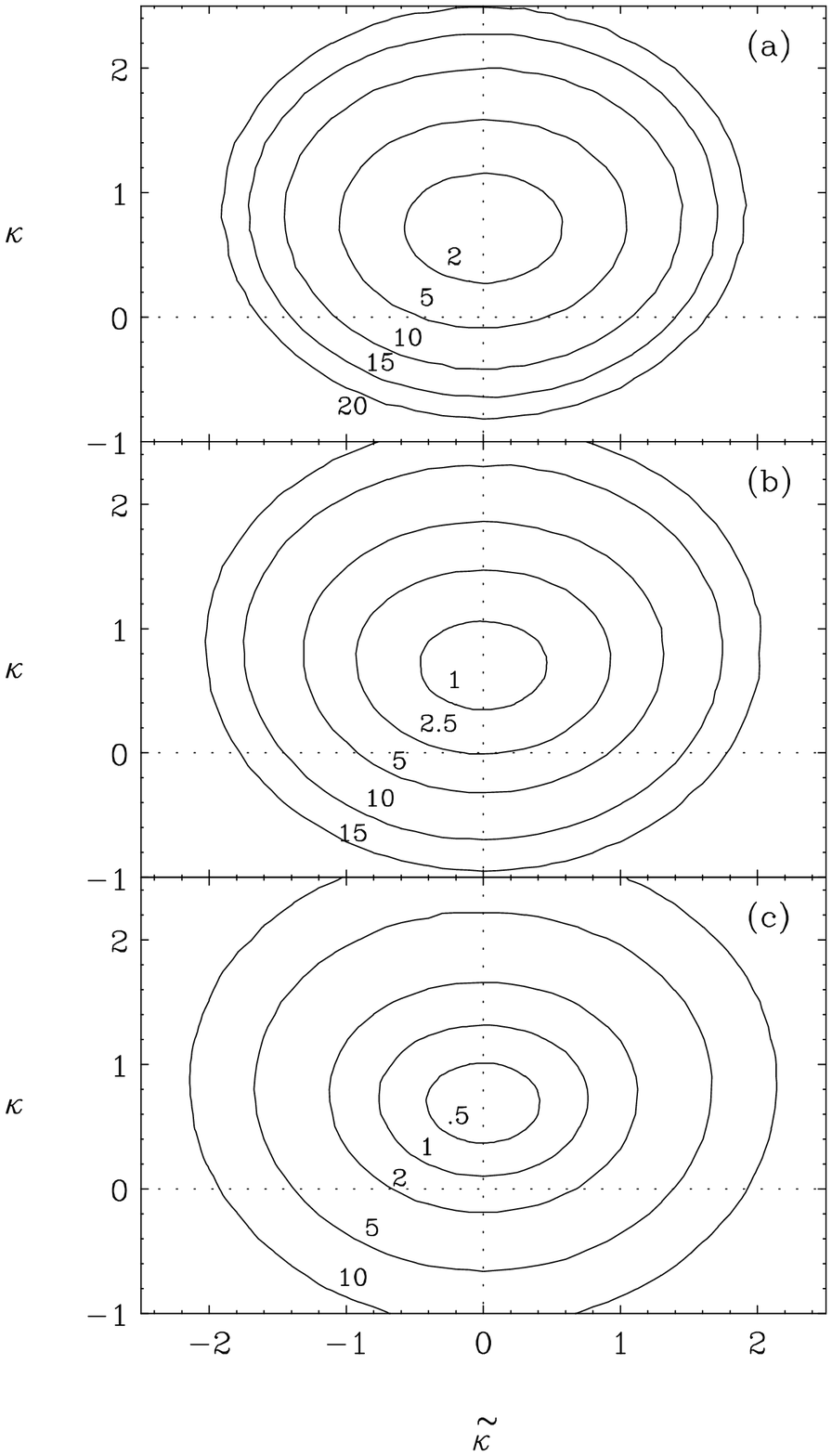}
\caption{}
\end{figure}

\begin{figure}
\centering
\leavevmode
\epsfysize=630pt
\epsfbox{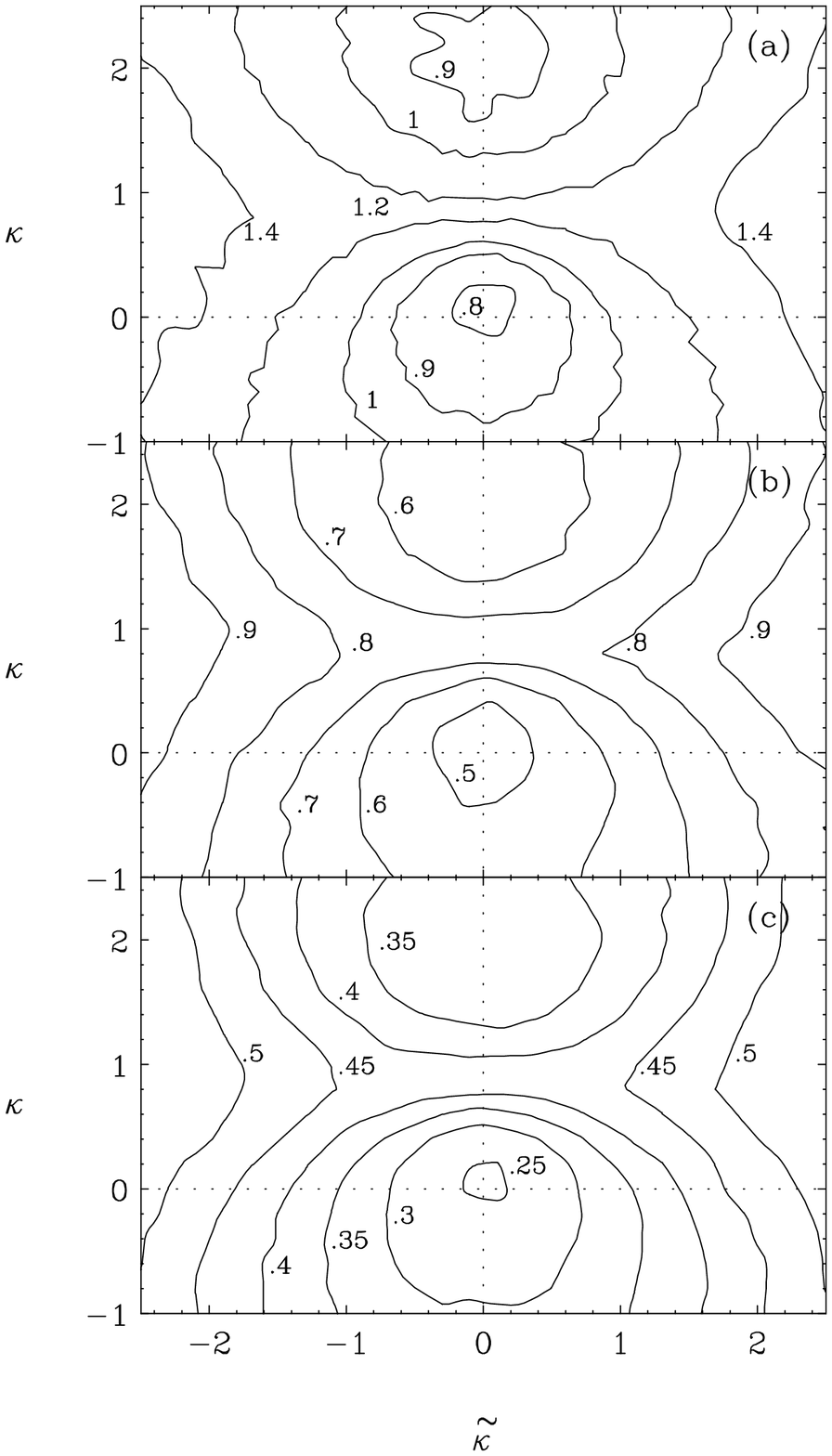}
\caption{}
\end{figure}

\begin{figure}
\centering
\leavevmode
\epsfysize=480pt
\epsfbox{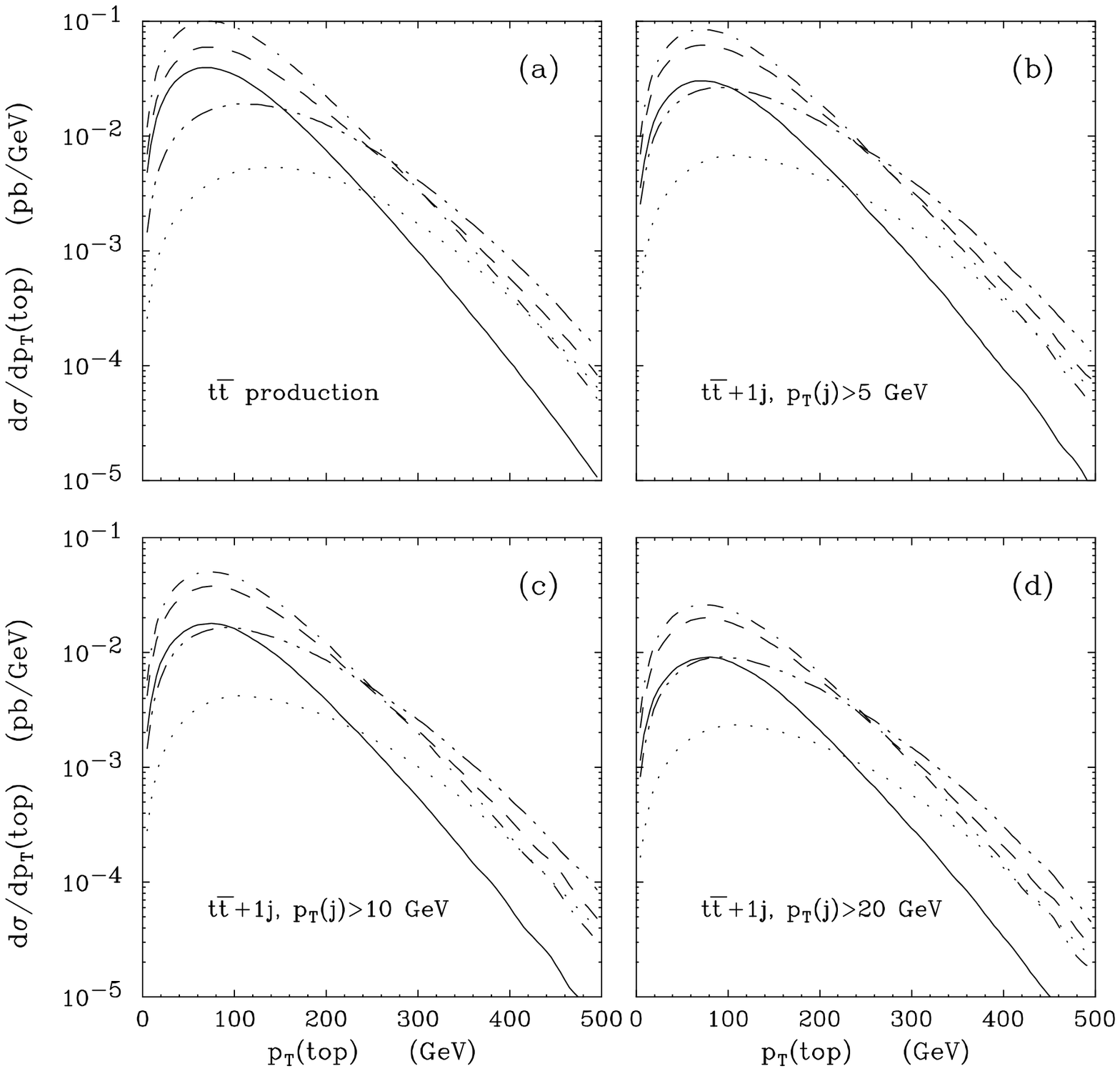}
\caption{}
\end{figure}

\begin{figure}
\centering
\leavevmode
\epsfysize=480pt
\epsfbox{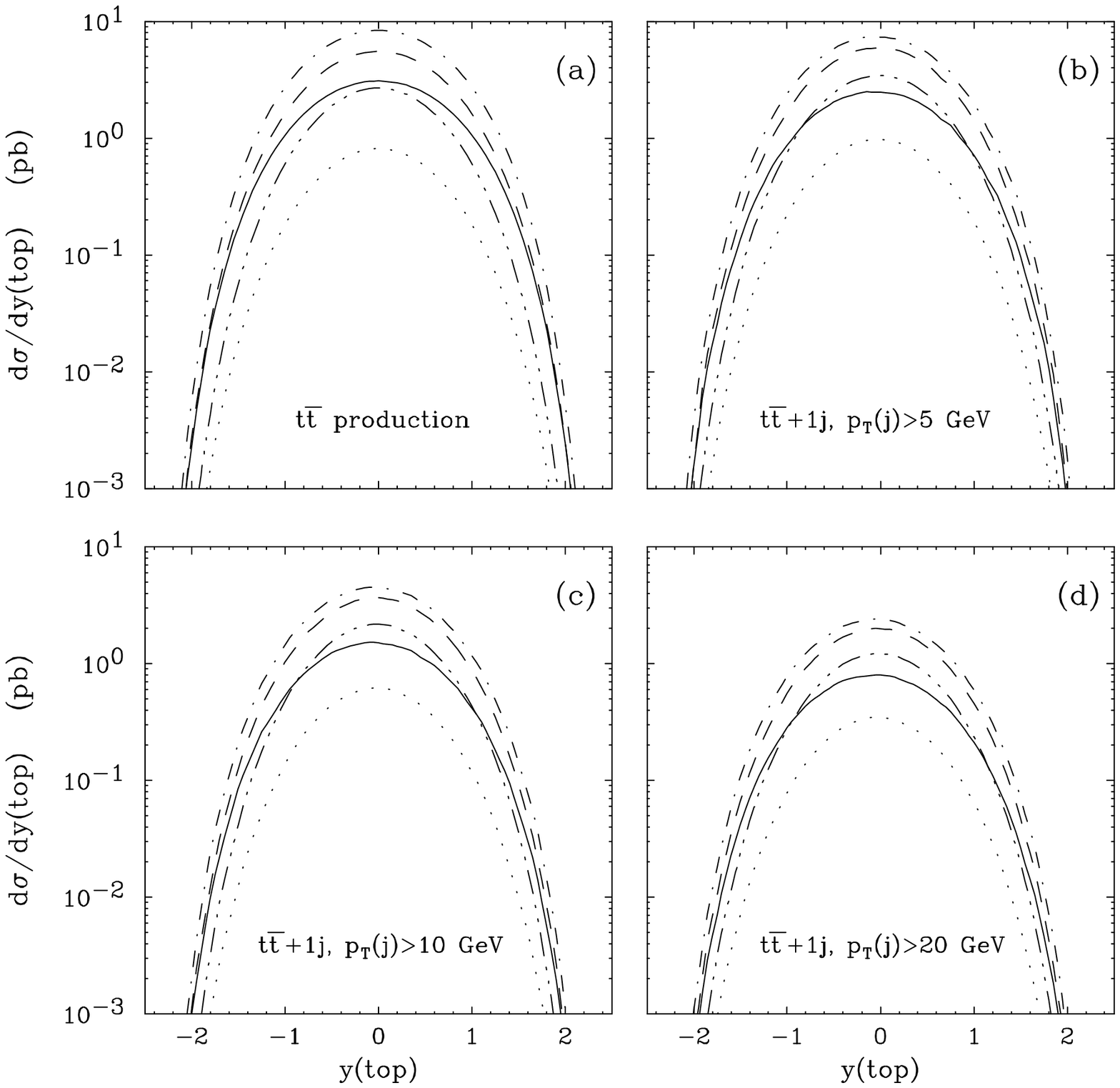}
\caption{}
\end{figure}

\begin{figure}
\centering
\leavevmode
\epsfysize=280pt
\epsfbox{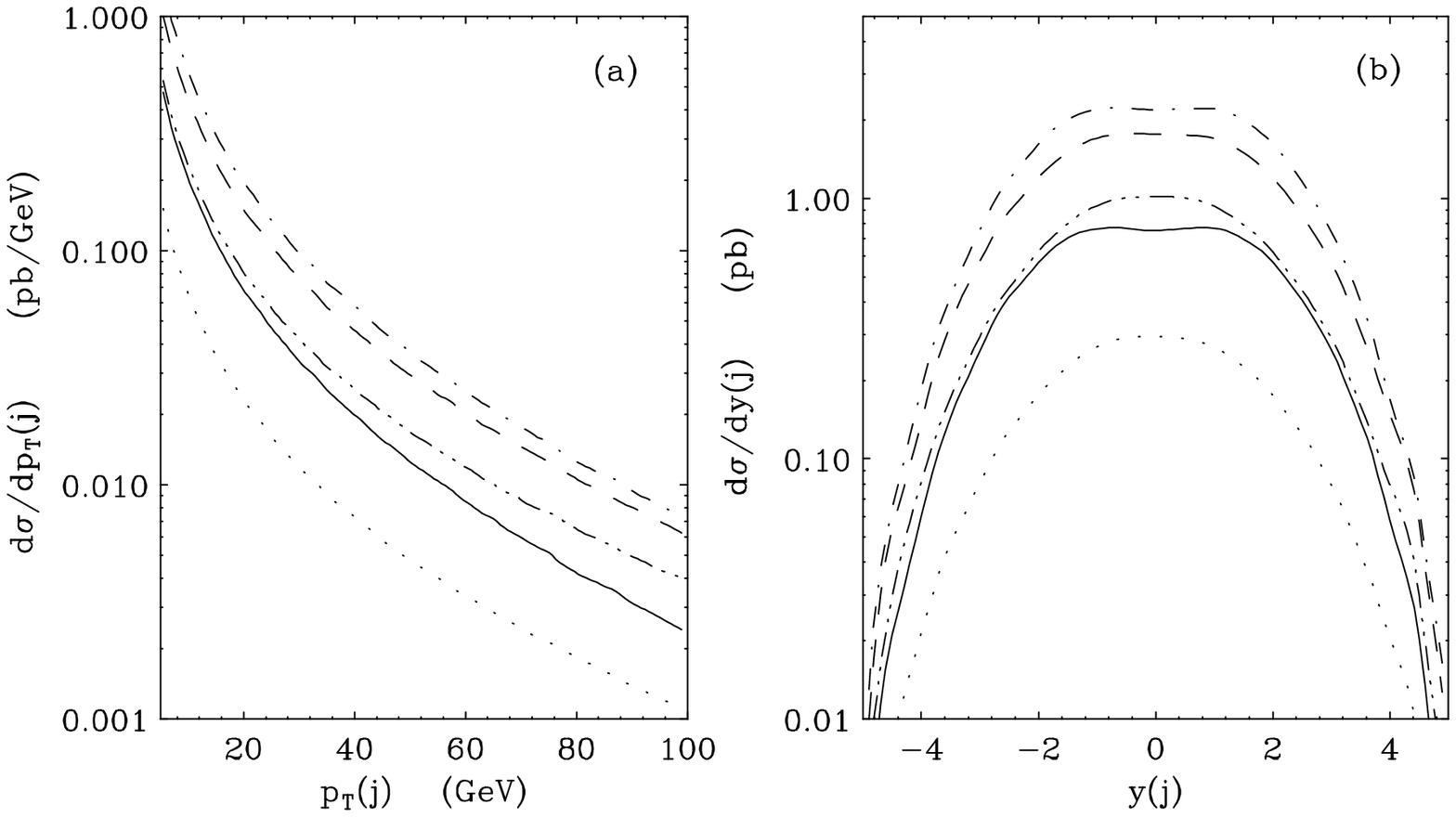}
\caption{}
\end{figure}

\end{document}